\errorstopmode
\input amssym.def
\input amssym.tex

% Page layout

\magnification=\magstephalf
\hsize=14.0 true cm
\vsize=19 true cm
\hoffset=1.0 true cm
\voffset=2.0 true cm

\abovedisplayskip=12pt plus 3pt minus 3pt
\belowdisplayskip=12pt plus 3pt minus 3pt
\parindent=1.0em

% Fonts

\font\sixrm=cmr6
\font\eightrm=cmr8
\font\ninerm=cmr9

\font\sixi=cmmi6
\font\eighti=cmmi8
\font\ninei=cmmi9

\font\sixsy=cmsy6
\font\eightsy=cmsy8
\font\ninesy=cmsy9

\font\sixbf=cmbx6
\font\eightbf=cmbx8
\font\ninebf=cmbx9

\font\eightit=cmti8
\font\nineit=cmti9

\font\eightsl=cmsl8
\font\ninesl=cmsl9

\font\sixss=cmss8 at 8 true pt
\font\sevenss=cmss9 at 9 true pt
\font\eightss=cmss8
\font\niness=cmss9
\font\tenss=cmss10

\font\sixmib=cmmib6
\font\sevenmib=cmmib7
\font\eightmib=cmmib8
\font\ninemib=cmmib9
\font\tenmib=cmmib10

 at 12 true pt
 at 12 true pt
\font\bigrm=cmr10 at 12 true pt
 at 12 true pt
 at 12 true pt

 at 16 true pt
%\font\Bigsy=cmsy12 at 16 true pt
%\font\Bigex=cmex12 at 16 true pt
 at 16 true pt
\font\Bigrm=cmr12 at 16 true pt
 at 16 true pt
 at 16 true pt

\catcode`@=11
\newfam\ssfam
\newfam\mibfam

\def\tenpoint{\def\rm{\fam0\tenrm}%
    \textfont0=\tenrm \scriptfont0=\sevenrm \scriptscriptfont0=\fiverm
    \textfont1=\teni  \scriptfont1=\seveni  \scriptscriptfont1=\fivei
    \textfont2=\tensy \scriptfont2=\sevensy \scriptscriptfont2=\fivesy
    \textfont3=\tenex \scriptfont3=\tenex   \scriptscriptfont3=\tenex
    \textfont\itfam=\tenit                  \def\it{\fam\itfam\tenit}%
    \textfont\slfam=\tensl                  \def\sl{\fam\slfam\tensl}%
    \textfont\bffam=\tenbf \scriptfont\bffam=\sevenbf
                           \scriptscriptfont\bffam=\fivebf
                           \def\bf{\fam\bffam\tenbf}%
    \textfont\ssfam=\tenss \scriptfont\ssfam=\sevenss
                           \scriptscriptfont\ssfam=\sevenss
                           \def\ss{\fam\ssfam\tenss}%
    \textfont\mibfam=\tenmib \scriptfont\mibfam=\sevenmib
                             \scriptscriptfont\mibfam=\sevenmib
                             \def\mib{\fam\mibfam\tenmib}%
    \normalbaselineskip=13pt
    \setbox\strutbox=\hbox{\vrule height8.5pt depth3.5pt width0pt}%
    \let\big=\tenbig
    \normalbaselines\rm}

\def\ninepoint{\def\rm{\fam0\ninerm}%
    \textfont0=\ninerm      \scriptfont0=\sixrm
                            \scriptscriptfont0=\fiverm
    \textfont1=\ninei       \scriptfont1=\sixi
                            \scriptscriptfont1=\fivei
    \textfont2=\ninesy      \scriptfont2=\sixsy
                            \scriptscriptfont2=\fivesy
    \textfont3=\tenex       \scriptfont3=\tenex
                            \scriptscriptfont3=\tenex
    \textfont\itfam=\nineit \def\it{\fam\itfam\nineit}%
    \textfont\slfam=\ninesl \def\sl{\fam\slfam\ninesl}%
    \textfont\bffam=\ninebf \scriptfont\bffam=\sixbf
                            \scriptscriptfont\bffam=\fivebf
                            \def\bf{\fam\bffam\ninebf}%
    \textfont\ssfam=\niness \scriptfont\ssfam=\sixss
                            \scriptscriptfont\ssfam=\sixss
                            \def\ss{\fam\ssfam\niness}%
    \textfont\mibfam=\ninemib \scriptfont\mibfam=\sixmib
                            \scriptscriptfont\mibfam=\sixmib
                            \def\mib{\fam\mibfam\ninemib}%
    \normalbaselineskip=12pt
    \setbox\strutbox=\hbox{\vrule height8.0pt depth3.0pt width0pt}%
    \let\big=\ninebig
    \normalbaselines\rm}

\def\eightpoint{\def\rm{\fam0\eightrm}%
    \textfont0=\eightrm      \scriptfont0=\sixrm
                             \scriptscriptfont0=\fiverm
    \textfont1=\eighti       \scriptfont1=\sixi
                             \scriptscriptfont1=\fivei
    \textfont2=\eightsy      \scriptfont2=\sixsy
                             \scriptscriptfont2=\fivesy
    \textfont3=\tenex        \scriptfont3=\tenex
                             \scriptscriptfont3=\tenex
    \textfont\itfam=\eightit \def\it{\fam\itfam\eightit}%
    \textfont\slfam=\eightsl \def\sl{\fam\slfam\eightsl}%
    \textfont\bffam=\eightbf \scriptfont\bffam=\sixbf
                             \scriptscriptfont\bffam=\fivebf
                             \def\bf{\fam\bffam\eightbf}%
    \textfont\ssfam=\eightss \scriptfont\ssfam=\sixss
                             \scriptscriptfont\ssfam=\sixss
                             \def\ss{\fam\ssfam\eightss}%
    \textfont\mibfam=\eightmib \scriptfont\mibfam=\sixmib
                             \scriptscriptfont\mibfam=\sixmib
                             \def\mib{\fam\mibfam\eightmib}%
    \normalbaselineskip=10pt
    \setbox\strutbox=\hbox{\vrule height7.0pt depth2.0pt width0pt}%
    \let\big=\eightbig
    \normalbaselines\rm}

\def\tenbig#1{{\hbox{$\left#1\vbox to8.5pt{}\right.\n@space$}}}
\def\ninebig#1{{\hbox{$\textfont0=\tenrm\textfont2=\tensy
                       \left#1\vbox to7.25pt{}\right.\n@space$}}}
\def\eightbig#1{{\hbox{$\textfont0=\ninerm\textfont2=\ninesy
                       \left#1\vbox to6.5pt{}\right.\n@space$}}}

\font\sectionfont=cmbx10
\font\subsectionfont=cmti10

\def\figurecaptionfont{\ninepoint}
\def\tablecaptionfont{\ninepoint}
\def\footnotefont{\eightpoint}

% New count registers

\newcount\equationno
\newcount\bibitemno
\newcount\figureno
\newcount\tableno

\equationno=0
\bibitemno=0
\figureno=0
\tableno=0
%\advance\pageno by -1

% Footline

\footline={\ifnum\pageno=0{\hfil}\else
{\hss\rm\the\pageno\hss}\fi}

% Section macro

\def\section #1. #2 \par
{\vskip0pt plus .10\vsize\penalty-100 \vskip0pt plus-.10\vsize
\vskip 1.6 true cm plus 0.2 true cm minus 0.2 true cm
\global\def\equationlabel{#1}
\global\equationno=0
\leftline{\sectionfont #1. #2}\par
\immediate\write\terminal{Section #1. #2}
\vskip 0.7 true cm plus 0.1 true cm minus 0.1 true cm
\noindent}

% Subsection macro

\def\subsection #1 \par
{\vskip0pt plus 0.8 true cm\penalty-50 \vskip0pt plus-0.8 true cm
\vskip2.5ex plus 0.1ex minus 0.1ex
\leftline{\subsectionfont #1}\par
\immediate\write\terminal{Subsection #1}
\vskip1.0ex plus 0.1ex minus 0.1ex
\noindent}

% Appendix macro

\def\appendix #1. #2 \par
{\vskip0pt plus .20\vsize\penalty-100 \vskip0pt plus-.20\vsize
\vskip 1.6 true cm plus 0.2 true cm minus 0.2 true cm
\global\def\equationlabel{\hbox{\rm#1}}
\global\equationno=0
\leftline{\sectionfont Appendix #1. #2}\par
\immediate\write\terminal{Appendix #1. #2}
\vskip 0.7 true cm plus 0.1 true cm minus 0.1 true cm
\noindent}

%\def\appendix #1. #2 \par
%{\vskip0pt plus .20\vsize\penalty-100 \vskip0pt plus-.20\vsize
%\vskip 1.6 true cm plus 0.2 true cm minus 0.2 true cm
%\global\def\equationlabel{\hbox{\rm#1}}
%\global\equationno=0
%\leftline{\sectionfont Appendix #1. #2}\par
%\immediate\write\terminal{Appendix #1. #2}
%\vskip 0.7 true cm plus 0.1 true cm minus 0.1 true cm
%\noindent}

% Displayed equations

\def\equation#1{$$\displaylines{\qquad #1}$$}
\def\enum{\global\advance\equationno by 1
\hfill\llap{{\rm(\equationlabel.\the\equationno)}}}
\def\noenum{\hfill}
\def\next#1{\cr\noalign{\vskip#1}\qquad}

% Bibliography macro, references

\def\ifundefined#1{\expandafter\ifx\csname#1\endcsname\relax}

\def\ref#1{\ifundefined{#1}?\immediate\write\terminal{unknown reference
on page \the\pageno}\else\csname#1\endcsname\fi}

\newwrite\terminal
\newwrite\bibitemlist

\def\bibitem#1#2\par{\global\advance\bibitemno by 1
\immediate\write\bibitemlist{\string\def
\expandafter\string\csname#1\endcsname
{\the\bibitemno}}
\item{[\the\bibitemno]}#2\par}

\def\beginbibliography{
\vskip0pt plus .15\vsize\penalty-100 \vskip0pt plus-.15\vsize
\vskip 1.2 true cm plus 0.2 true cm minus 0.2 true cm
\leftline{\sectionfont References}\par
\immediate\write\terminal{References}
\immediate\openout\bibitemlist=biblist
\frenchspacing\parindent=1.8em
\vskip 0.5 true cm plus 0.1 true cm minus 0.1 true cm}

\def\endbibliography{
\immediate\closeout\bibitemlist
\nonfrenchspacing\parindent=1.0em}

% Figure and table captions

\def\figurecaption#1{\global\advance\figureno by 1
\narrower\figurecaptionfont
Fig.~\the\figureno. #1}

\def\tablecaption#1{\global\advance\tableno by 1
\vbox to 0.25 true cm { }
\centerline{\tablecaptionfont%
Table~\the\tableno. #1}
\vskip-0.4 true cm}

\def\thintablerule{\hrule height0.4pt}

\tenpoint

\immediate\openin\bibitemlist=biblist
\ifeof\bibitemlist\immediate\closein\bibitemlist
\else\immediate\closein\bibitemlist
\input biblist \fi

% current year and month

\def\thismonth{\ifcase\month\or
January\or February\or March\or April\or May\or June\or
July\or August\or September\or October\or November\or December\fi}

\input epsf
\epsfclipon

% Definitions and abbreviations

% Roman letters in math formulae

\def\rmd{{\rm d}}

\def\rme{{\rm e}}
\def\rmO{{\rm O}}

% Real and integer numbers

\def\gz{{\Bbb Z}}

% Special relations and symbols

\def\proof{\noindent{\sl Proof:}\kern0.6em}

\def\frac#1#2{\hbox{$#1\over#2$}}
\def\dual{\mathstrut^*\kern-0.1em}
\def\mod{\;\hbox{\rm mod}\;}

\def\lvec#1{\setbox0=\hbox{$#1$}
    \setbox1=\hbox{$\scriptstyle\leftarrow$}
    #1\kern-\wd0\smash{
    \raise\ht0\hbox{$\raise1pt\hbox{$\scriptstyle\leftarrow$}$}}
    \kern-\wd1\kern\wd0}
\def\rvec#1{\setbox0=\hbox{$#1$}
    \setbox1=\hbox{$\scriptstyle\rightarrow$}
    #1\kern-\wd0\smash{
    \raise\ht0\hbox{$\raise1pt\hbox{$\scriptstyle\rightarrow$}$}}
    \kern-\wd1\kern\wd0}
\def\slash#1{\setbox0=\hbox{$#1$}\setbox1=\hbox{$\kern1pt/$}
    #1\kern-\wd0\kern1pt/\kern-\wd1\kern\wd0}

% Lattice derivatives

\def\nab#1{{\nabla_{#1}}}
\def\nabstar#1{{\nabla\kern0.5pt\smash{\raise 4.5pt\hbox{$\ast$}}
               \kern-5.5pt_{#1}}}
\def\drv#1{{\partial_{#1}}}
\def\drvstar#1{{\partial\kern0.5pt\smash{\raise 4.5pt\hbox{$\ast$}}
               \kern-6.0pt_{#1}}}

\def\ldrvstar#1{{\lvec{\,\partial}\kern-0.5pt\smash{\raise 4.5pt\hbox{$\ast$}}
               \kern-5.0pt_{#1}}}

% Units

% Constants

% Fields

\def\psibar{\overline{\psi}}
\def\zetabar{\overline{\zeta}}

\def\bfield{{\cal O}}

% Dirac matrices

\def\dirac#1{\gamma_{#1}}
\def\diracstar#1#2{
    \setbox0=\hbox{$\gamma$}\setbox1=\hbox{$\gamma_{#1}$}
    \gamma_{#1}\kern-\wd1\kern\wd0
    \smash{\raise4.5pt\hbox{$\scriptstyle#2$}}}

% Gauge group

\def\Ad{{\rm Ad}\kern0.1em}

% Action

\def\Sf{S_{\rm F}}

% Parameters and abbreviations

\def\Dw{D_{\rm w}}
\def\Dm{D_m}
\def\hDm{Q_m}
\def\U{U}
\def\Utilde{U\tilde{\phantom{a}}}
\def\Ahat{\hat{A}}
\def\Bhat{\hat{B}}
\def\abar{\bar{a}}

%
%\vbox{\vskip0.0cm}
\rightline{CERN-PH-TH/2006-052}

\vskip 1.5cm 
\centerline{\Bigrm
The Schr\"odinger functional in lattice QCD
}
\vskip 0.3 true cm
\centerline{\Bigrm 
with exact chiral symmetry
}
\vskip 0.6 true cm
\centerline{\bigrm Martin L\"uscher}
\vskip1ex
\centerline{\it CERN, Physics Department, TH Division}
\centerline{\it CH-1211 Geneva 23, Switzerland}
\vskip 0.8 true cm
\thintablerule
\vskip 2.0ex
\ninepoint
\leftline{\bf Abstract}
\vskip 1.0ex\noindent
Similarly to the interaction lagrangian, the possible
boundary conditions in quantum field theories on space-time
manifolds with boundaries are strongly constrained by the
symmetry and scaling properties of the theory.
Based on this general insight, a lattice formulation
of the QCD Schr\"odinger functional is proposed for the case
where the lattice Dirac operator in the bulk of the lattice
coincides with the Neuberger--Dirac operator.
The construction 
satisfies all basic requirements (locality, symmetries, hermiticity)
and is suitable for numerical simulations.
\vskip 2.0ex
\thintablerule

\tenpoint

\vskip-0.2cm

\section 1. Introduction

In numerical lattice QCD, the Schr\"odinger functional 
mainly serves as a probe in scaling studies of the theory close to the
continuum limit [\ref{SchroedingerGauge},\ref{SchroedingerFermion}].
The non-perturbative 
computation of renormalization factors, for example,
is an important case where the Schr\"odinger functional proved
to be very useful
(see ref.~[\ref{LesHouches}] for an introduction).

In the continuum limit the Schr\"odinger functional
is obtained by restricting the 
time coordinate $x_0$ to a finite range, $0\leq x_0\leq T$,
and by imposing Dirichlet boundary conditions on the fields
at $x_0=0$ and $x_0=T$.
The boundary conditions break chiral symmetry, 
but the chiral Ward identities remain valid away from the boundaries
and thus forbid the mixing of operator insertions belonging to different
chiral multiplets, for example,
as is the case in the absence of the boundaries.

The symmetry properties of the Schr\"odinger functional
in lattice QCD should ideally be the same if
the chosen lattice Dirac operator  
preserves chiral symmetry on the infinite lattice
via the Ginsparg--Wilson relation
[\ref{GinspargWilson}--\ref{ExactChSy}].
However, in the presence of the boundaries, 
the Dirac operator and the Ginsparg--Wilson
relation must both be modified.
Chiral symmetry would otherwise
not be broken and the lattice Schr\"odinger functional could
not possibly have the correct continuum limit.
The modifications must evidently be local and 
be linked to the boundary conditions,
but without further insight it is difficult to say 
how to proceed from here.

A theoretically attractive possibility, recently studied
by Taniguchi [\ref{OrbiTaniguchi}], is
to define the Schr\"odinger functional through an orbifold projection.
The boundary conditions are 
determined by the $\gz_2$ orbifold symmetry in this case, which
is taken to be the product of a time reflection and a chiral rotation.
As it turns out, however,
the latter leads to some technical difficulties on the lattice
and one ends up with a lattice Dirac operator whose determinant
has a non-removable phase.
Moreover,
the construction becomes rather artificial
when the masses of the quarks do not vanish%
$\,$\footnote{$\dagger$}{\footnotefont%
All these problems disappear
if chirally rotated boundary
conditions are adopted,
as suggested by Sint [\ref{OrbiSint}].
The goal here, however, is
to define the Schr\"odinger functional with the standard~parity\-conserving 
boundary conditions [\ref{SchroedingerFermion}].}.

To a large extent, the solution proposed in this paper is based on 
universality considerations. Stated somewhat superficially,
the idea is that Schr\"odinger functional boundary conditions
do not require any fine-tuning and will thus be satisfied automatically
in the continuum limit, as long as the lattice theory in 
the presence of the boundaries  
respects locality and the obvious lattice symmetries.
The problem then reduces to finding a lattice
Dirac operator that has these properties and additionally
satisfies the Ginsparg--Wilson relation away from the 
lattice boundaries. 

Starting from the Neuberger--Dirac operator in infinite volume 
[\ref{NeubergerDirac}], 
such operators are not too difficult to construct.
It may be useful, however, to first recall some basic facts on
the quark sector of 
the Schr\"odinger functional (sect.~2) and to address the 
issue of universality in some detail (sect.~3). 
In sect.~4, Ginsparg--Wilson fermions in one dimension are briefly 
discussed and it is then only a small step to write down 
an acceptable lattice Dirac operator in four dimensions
(sect.~5).

\section 2. Quark fields and Dirac operator in the continuum theory

In the following, it will be assumed that the reader is familiar
with the Schr\"odinger functional in QCD
[\ref{SchroedingerGauge},\ref{SchroedingerFermion}].
Most of the time the SU(3) gauge field will play
a spectator r\^ole and its presence can be largely ignored.
In this section some properties of the quark sector 
in the continuum theory are highlighted.
This serves partly to introduce the subject and partly
to guide the construction of the lattice theory in the later
sections.

\subsection 2.1 Boundary conditions

It suffices to consider a single quark with mass $m$,
since all formulae and comments trivially extend to the 
case of several quarks with arbitrary masses.
The quark and antiquark fields $\psi(x)$
and $\psibar(x)$ are defined at all times $x_0\in[0,T]$
and are required to satisfy$\,$\footnote{$\ddag$}{\footnotefont%
The notation is the one commonly used in lattice QCD.
In particular, the space-time metric is euclidean and 
the Dirac matrices $\dirac{\mu}$, $\mu=0,\ldots,3$, are taken
to be hermitian. The fifth Dirac matrix, 
$\dirac{5}=\dirac{0}\dirac{1}\dirac{2}\dirac{3}$, is then also 
hermitian. Where appropriate, the Einstein summation convention
is applied.}
\equation{
  P_{+} \psi(x)=\psibar(x)P_{-}=0\quad\hbox{at}\quad x_0=0,
  \enum
  \next{2.5ex}
  P_{-} \psi(x)=\psibar(x)P_{+}=0\quad\hbox{at}\quad x_0=T,
  \enum
}
where $P_{\pm}=\frac{1}{2}(1\pm\dirac{0})$.
These boundary conditions are invariant under 
space rotations, parity, time reflections ($x_0\to T-x_0$)
and charge conjugation. 

The quark action in the presence of a gauge field
$A_{\mu}(x)$ is then given by
\equation{
  \Sf=\int_0^T\rmd x_0\int\rmd^3{\mib x}\,
  \psibar(x)\Dm\psi(x),
  \enum
  \next{2ex}
  \Dm=\dirac{\mu}D_{\mu}+m,\qquad
  D_{\mu}=\partial_{\mu}+A_{\mu}.
  \enum
}
Usually the Schr\"odinger functional is set up
with inhomogeneous boundary conditions, where the boundary values
serve as sources for the quark field at $x_0=0$ and $x_0=T$ 
[\ref{SchroedingerFermion}].
One may, however, just as well adopt homogeneous boundary conditions,
as is done here, 
and introduce the boundary quark fields
directly through
\equation{
   \zeta({\mib x})=P_{-}\psi(x),
   \qquad \zetabar({\mib x})=\psibar(x)P_{+}
   \quad\hbox{at}\quad x_0=0,
   \enum
   \next{2ex}
   \zeta'({\mib x})=P_{+}\psi(x), 
   \qquad \zetabar\vphantom{\zeta}'({\mib x})=\psibar(x)P_{-}
   \quad\hbox{at}\quad x_0=T.
   \enum
}
These are just the non-zero Dirac components of the quark fields at the 
boundaries.

\subsection 2.2 Spectrum of the Dirac operator

Let $\cal H$ be the linear space of all smooth quark fields
that satisfy the required boundary conditions.
With respect to the natural scalar product in this space,
\equation{
  \left(\psi,\chi\right)=
  \int_0^T\rmd x_0\int\rmd^3{\mib x}\,
  \psi(x)^{\dagger}\chi(x),
  \enum
}
the operator $\hDm=\dirac{5}\Dm$ is easily shown to be hermitian.
Since $\hDm$ is also elliptic,
it follows that this operator has a complete orthonormal
set of eigenfunctions $v_n\in\cal H$, labelled by an index
$n=0,1,2,\ldots$,
with real eigenvalues $\lambda_n$.

For non-negative quark masses $m$, the eigenvalues are bounded 
from below by
\equation{
   \lambda_n^2\geq m^2+\mu^2,
   \enum
}
where $\mu$ denotes the spectral gap at $m=0$. This is a straightforward
consequence of the identity
\equation{
   \lambda_n^2=\left\|\hDm v_n\right\|^2=
   m^2\left\|v_n\right\|^2+
   \left\|\left.\hDm\right|_{m=0}v_n\right\|^2
   \noenum
   \next{2.5ex}
   {\phantom{\lambda_n^2=}}
   +m\int_{x_0=0}\rmd^3{\mib x}\,v_n(x)^{\dagger}v_n(x)
   +m\int_{x_0=T}\rmd^3{\mib x}\,v_n(x)^{\dagger}v_n(x).
   \enum
}
Moreover, using the fact that 
the current $v_n(x)^{\dagger}\dirac{\mu}v_n(x)$ is conserved at 
$m=\lambda_n=0$,
it is possible to show that the massless Dirac operator has no zero modes.

Note that the bound (2.8) 
becomes invalid at $m<0$. The Dirac operator
has eigenmodes 
localized at the boundaries in this case, with eigenvalues that
decrease exponentially at large $T$. These are just the well-known
domain wall fermion modes. The fact that the sign of the quark
mass matters is a consequence of the breaking of chiral symmetry
through the boundary conditions. This is also why the eigenvalues
at non-zero quark masses are not simply related to those at vanishing
mass.

\subsection 2.3 Determinant of the Dirac operator

Since there are positive and negative eigenvalues $\lambda_n$,
the determinant, $\det \hDm$, may have a non-trivial phase.
However, this is not the case if
the standard lattice regularization 
is employed [\ref{SchroedingerFermion}].
Other regularizations may give different results,
but the universality of the continuum limit 
implies that the determinant must always
be real after removal of the ultraviolet cutoff, 
up to some local terms perhaps.
These can always be ``renormalized away'',
and the requirement of a real determinant 
then becomes part of the definition of the theory.

At non-negative quark masses, the quark determinant
actually has a definite sign in this case,
because the eigenvalues of $\hDm$ never pass through zero.
In the functional integral,
the quark determinant may thus be 
replaced by its absolute value.

\subsection 2.4 Chiral symmetry properties of the quark propagator at $m=0$

As a differential operator, the massless Dirac operator 
anticommutes with $\dirac{5}$ and the local chiral Ward idenitites 
are thus not affected by the presence of the boundaries.
The quark propagator $S(x,y)$, on the other hand, 
transforms according to 
\equation{
  \dirac{5}S(x,y)+S(x,y)\dirac{5}=
  \noenum
  \next{2.5ex}
  \qquad
  \int_{z_0=0}\rmd^3{\mib z}\,S(x,z)\dirac{5}S(z,y)+
  \int_{z_0=T}\rmd^3{\mib z}\,S(x,z)\dirac{5}S(z,y).
  \enum
}  
This equation is easily established by noting that, as a function of $x$,
the expression on the left solves the 
homogenous Dirac equation.
The other side of the equation is then obtained by reconstructing 
the solution from its boundary values at $x_0=0$ and $x_0=T$.

In QCD with more than one massless quark,
eq.~(2.10) implies the non-singlet chiral Ward identity
\equation{
  \left\langle
  \lambda^a\dirac{5}\psi(x)\psibar(y)+\psi(x)\psibar(y)\lambda^a\dirac{5}
  \right\rangle_{\rm F}=
  \noenum
  \next{2.5ex}
  \qquad
  \int\rmd^3{\mib z}  
  \left\langle\psi(x)\psibar(y)
  \left\{\zetabar({\mib z})\lambda^a\dirac{5}\zeta({\mib z})+
         \zetabar\vphantom{\zeta}'({\mib z})\lambda^a\dirac{5}\zeta'({\mib z})
  \right\}
  \right\rangle_{\rm F},
  \enum
}
where $\lambda^a$ is any traceless matrix in flavour space
and $\langle\ldots\rangle_{\rm F}$ denotes the expectation value
in the quark sector.
The breaking of chiral symmetry through
the boundary conditions is made explicit by this formula.
Purely from the symmetry point of view, it actually looks as if 
there were a unit mass term at the boundaries.

\section 3. Field theories on lattices with boundaries

In continuum field theories,
the notion of smoothness plays a central r\^ole. 
Differential operators can only act on smooth functions, for example, and 
boundary conditions really only make sense when imposed on 
such functions.

The situation in lattice field theory is quite different.
Strictly speaking, boundary conditions are
no longer imposed on the fields but instead are 
encoded in the lattice action.
As usual different lattice theories may have the same continuum limit,
where now the characterization of the latter must include a 
specification of the boundary conditions. 
An important point to note is that
the possible boundary conditions are strongly
constrained by the requirement of locality, the lattice symmetries
and by power-counting arguments.
There are then not many more universality classes 
than there are in the absence of boundaries.

The aim in this section is to make these remarks a bit more
concrete, so that they can be applied to the Schr\"odinger functional
in QCD. Eventually, the argumentation is based on 
Symanzik's work on the renormalization of quantum field theories
in the presence of boundaries 
[\ref{SchrodingerSymanzik}]
and also on the theory of
boundary critical phenomena in statistical mechanics
(see ref.~[\ref{Diehl}] for a review).

\subsection 3.1 Boundary conditions and the field equations

The concepts that will be developed in the following
are best introduced by considering a free scalar field
in the half-space $x_0\geq0$ with various boundary conditions 
at $x_0=0$.
A simple lattice formulation of this theory is obtained by choosing  
the lattice field $\phi(x)$ to reside on the sites
of a hypercubic lattice with spacing $a$.
The expression
\equation{
  S=a^4\sum_{x_0\geq a}\sum_{\mib x}
  \frac{1}{2}\left\{\drv{\mu}\phi(x)\drv{\mu}\phi(x)+
  m^2\phi(x)^2\right\}
  \enum
}
is then a possible choice of the lattice action,
where $\drv{\mu}$ denotes
the forward nearest-neighbour difference operator in direction $\mu$
and the mass $m$ is assumed to be positive. 
Note that the action depends on the field variables at time
$x_0=a,2a,3a,\ldots$ only. These are thus
the unconstrained dynamical degrees of freedom of the field
which are to be integrated over in the functional integral.

Starting from the action (3.1),
the propagator $\langle\phi(x)\phi(y)\rangle$ 
can easily be worked out in a time--momentum representation.
It then turns out that the propagator satisfies Neumann boundary
conditions in the continuum limit. 
On the other hand, after a slight modification of the action
by a boundary term,
\equation{
  S\to S+a^3\sum_{\mib x}{c\over2a}\left.\phi(x)^2\right|_{x_0=a},
  \enum
}
the calculation yields a propagator that
satisfies Dirichlet boundary conditions in the continuum limit,
for any fixed $c>0$ 
(the powers of $a$ are
such that $c$ is dimensionless). 

Some understanding of how a particular boundary condition arises
can be achieved
by considering the field equation
\equation{
  \langle\eta(x)\phi(y)\rangle=a^{-4}\delta_{xy},
  \qquad\eta(x)={\delta S\over\delta\phi(x)}.
  \enum
}
A short calculation yields
\equation{
  \eta(x)=\{-\drvstar{\mu}\drv{\mu}+m^2\}\phi(x)
  \quad\hbox{at}\quad x_0>a,
  \enum
}
where $\drvstar{\mu}$ denotes the backward nearest-neighbour 
difference operator. In the bulk of the lattice, the field equation 
thus coincides with the lattice Klein--Gordon equation.
At $x_0=a$, however, a different equation is obtained,
\equation{
  \eta(x)={c\over a^2}\phi(x)-{1\over a}\drv{0}\phi(x)+
  \{-\drvstar{k}\drv{k}+m^2\}\phi(x),
  \enum
}
which depends on the details of the action close to the boundary.
Formally, the first term in this expression dominates
in the continuum limit and the field equation at the boundary 
thus implies Dirichlet boundary conditions at $a=0$ if $c>0$. 
On the other hand, the second term dominates if $c=0$, which 
leads to Neumann boundary conditions in the continuum limit.

\subsection 3.2 Natural boundary conditions

The example considered in the previous subsection 
illustrates the fact that boundary conditions 
are a property of the continuum theory which
arises dynamically in the continuum limit. In particular,
they are not determined by the space of lattice fields
alone.
Another outcome is that some boundary conditions 
are obtained generically, while others 
(Neumann and mixed boundary conditions in the study case)
are unstable under perturbations of the lattice action.

In an interacting theory, 
and also in free theories with complicated
lattice actions,
the connection between the lattice field
equations and the boundary conditions in the continuum limit
may not be as transparent as suggested above.
It seems plausible, however, that the boundary conditions
are always of the form $\left.\bfield(x)\right|_{x_0=0}=0$,
where $\bfield(x)$ is a linear combination of 
local fields with the appropriate 
number of components and symmetry properties.

The boundary conditions that are
stable under perturbations of the lattice
theory, and which thus arise naturally,
then correspond to the fields with the lowest possible dimension.
Some tuning of the lattice theory will normally be required
in all other cases,
unless there are some symmetries that protect 
$\bfield(x)$ from mixing with
lower dimensional fields.

\subsection 3.3 Are Schr\"odinger functional boundary conditions natural?

Following Wilson [\ref{Wilson}], the quark fields are represented
on the lattice by Dirac spinors $\psi(x)$ that reside on the points
$x$ of the lattice. Since QCD is asymptotically free, the 
scaling dimension of the local fields in this theory
coincides with their engineering dimension (up to logarithms).
In particular, the fields of lowest dimension
that carry the quantum numbers of the quark fields 
are the quark fields themselves.

The boundary conditions at $x_0=0$, which
arise naturally in the continuum limit, 
are thus of the form $\left.B\psi(x)\right|_{x_0=0}=0$, 
where $B$ is a constant matrix in Dirac and colour space.
At $x_0=T$ the boundary conditions are linked to
those at $x_0=0$ by the time reflection symmetry and
charge conjugation then determines the 
boundary conditions for the antiquark field.

If the lattice theory is invariant under gauge transformations,
cubic rotations and parity, as will be the case in the following,
the boundary conditions must respect these symmetries$\,$%
\footnote{$\dagger$}{\footnotefont%
Gauge transformations include the gauge field variables at the boundaries
and are thus a spurion symmetry to some extent.
In the present context, this is of no importance, however,
since $B$ does not depend on the gauge field.}.
Moreover, $B$ cannot have maximal rank,
as otherwise the boundary conditions and the Dirac equation
at $0<x_0<T$
would imply a vanishing quark propagator.
Up to a normalization constant,
the only matrices that are compatible with all these conditions
then are $B=P_{+}$ and $B=P_{-}$.

Schr\"odinger functional
boundary conditions thus arise naturally and do not
require any particular adjustments of the lattice action.
There are two classes of lattice theories,
which are distinguished by the sign in the 
boundary condition $\left.P_{\pm}\psi(x)\right|_{x_0=0}=0$.
The difference matters if the quark mass 
does not vanish, but the sign may easily be determined,
in any given case,
by studying the free-quark propagator for example.

\section 4. Ginsparg--Wilson quarks in one dimension

In the presence of the boundaries,
an acceptable lattice Dirac operator
that satisfies the Ginsparg--Wilson relation in the bulk 
of the lattice still needs to be found.
The Wilson--Dirac operator provides  
a simple solution to this problem in one dimension.
This is a somewhat trivial case, but it gives important hints for the 
construction of the Dirac operator in higher dimensions.

\subsection 4.1 Lattice Dirac operator

For simplicity the gauge field is omitted in this section.
The Wilson--Dirac operator in one dimension then reads
\equation{
   D=\frac{1}{2}\left\{
   \dirac{0}(\drvstar{0}+\drv{0})-a\drvstar{0}\drv{0}
   \right\}=P_{+}\drvstar{0}-P_{-}\drv{0}.
   \enum
} 
In the absence of boundaries, 
this operator satisfies
the Ginsparg--Wilson relation
\equation{
  \dirac{5}D+D\dirac{5}=aD\dirac{5}D.
  \enum
}
Moreover, from the definition (4.1) it is 
immediate that $\dirac{5}D$ is hermitian and that
the quark propagator is given by
\equation{
  S(x_0,y_0)=\theta(x_0\geq y_0)P_{+}+\theta(x_0\leq y_0)P_{-},
  \enum
}
where $\theta(\ast)$ is equal to $1$
if the logical condition in brackets is true and $0$ otherwise.

Following the standard treatment [\ref{SchroedingerFermion}],
the dynamical degrees of freedom of the quark fields in the
presence of the boundaries at $x_0=0$ and $x_0=T$ are taken 
to be their components at $x_0=a,2a,\ldots T-a$. 
It is convenient
to assume that the fields are defined at all other times as well, but
that they are equal to zero there.

The Dirac operator in the presence of the boundaries
(which is also denoted
by $D$) maps this space of quark fields into itself.
At $x_0=a,2a,\ldots,T-a$,
its action is given by eq.~(4.1), while at $x_0\leq0$ and $x_0\geq T$
the target field is set to zero.
With respect to the scalar product
\equation{
   \left(\psi,\chi\right)=a\sum_{x_0=a}^{T-a}\psi(x_0)^{\dagger}\chi(x_0),
   \enum
} 
the operator $\dirac{5}D$ is then again hermitian. It is also easy to show 
that $D$ has
no zero modes and that the associated propagator
coincides with the propagator (4.3) on the infinite lattice, 
at all times in the range $0<x_0,y_0<T$. 
In particular, the correct
Schr\"odinger functional boundary conditions
are obtained in the continuum limit.

\subsection 4.2 Modified Ginsparg--Wilson relation

As already pointed out in sect.~1,
the lattice Dirac operator is not
expected to satisfy the Ginsparg--Wilson relation
in the presence of the boundaries, for general reasons.
On the finite lattice, the operator $D$ introduced above
in fact satisfies the modified relation
\equation{
  \dirac{5}D+D\dirac{5}=aD\dirac{5}D+\dirac{5}P,
  \enum
}
where $P$ is a local operator given by
\equation{
  P\psi(x_0)={1\over a}\left\{
  \delta_{x_0a}P_{-}\psi(a)+\delta_{x_0\,T-a}P_{+}\psi(T-a)
  \right\}.
  \enum
}
Equation (4.5) is a straightforward consequence
of the precise definition of $D$ in the presence of the boundaries.
In particular, the boundary term $\dirac{5}P$ 
is obtained when the product $D\dirac{5}D$ 
is worked out at $x_0=a$ and $x_0=T-a$.

By multiplication of the modified Ginsparg--Wilson relation (4.5)
from both sides with the quark propagator, it follows that 
\equation{
  \dirac{5}S(x_0,y_0)+S(x_0,y_0)\dirac{5}=
  \dirac{5}\delta_{x_0y_0}+
  \noenum
  \next{2.5ex}
  \qquad
  S(x_0,a)\dirac{5}P_{-}S(a,y_0)+
  S(x_0,T-a)\dirac{5}P_{+}S(T-a,y_0).
  \enum
}  
In the continuum limit, 
the projectors $P_{\pm}$ on the right-hand side
of this equation can be dropped 
in view of the boundary conditions satisfied by the propagator.
The relation then reduces to the one-dimensional 
form of the chiral Ward identity (2.10).
This shows that the presence of the boundary 
term in the modified Ginsparg--Wilson relation (4.5) is directly 
linked to the 
breaking of chiral symmetry in the continuum theory by 
the boundary conditions.

\section 5. Lattice Dirac operator in four dimensions

Since Schr\"odinger functional boundary conditions arise naturally,
the choice of the lattice Dirac operator is not critical 
and there are probably many viable constructions.
The operator proposed here is a simple modification of the 
Neuberger--Dirac operator in infinite volume.

\subsection 5.1 Definition

As before the theory is first set up on the infinite lattice.
The quark fields are thus assumed to be defined 
at all sites of the lattice. Although the gauge field
continues to play a spectator r\^ole, it is now 
included in the formulae.
The Wilson--Dirac operator is then given by [\ref{Wilson}]
\equation{
  \Dw=\frac{1}{2}\left\{\dirac{\mu}(\nabstar{\mu}+\nab{\mu})-
  a\nabstar{\mu}\nab{\mu}\right\},
  \enum
}
where $\nab{\mu}$ and $\nabstar{\mu}$ denote the gauge-covariant
forward and backward difference operators.
As already mentioned, the starting point in this section is  
the Neuberger--Dirac operator [\ref{NeubergerDirac}]
\equation{
  D={1\over\abar}\bigl\{1-A\left(A^{\dagger}A\right)^{-1/2}\bigr\},
  \enum
  \next{2ex}
  A=1+s-a\Dw,\qquad
  \abar={a\over 1+s}.
  \enum
}
The parameter $s$ in this formula
allows for some optimization and is 
only relevant in the context of numerical simulations.
In practice, it is normally set to a fixed value in
the range $0\leq s\leq1/2$.

In the presence of the boundaries at $x_0=0$ and $x_0=T$, 
the dynamical degrees of freedom of the quark fields reside on the 
lattice sites at time $x_0=a,2a,\ldots,T-a$. It is
again convenient to assume that the fields are defined
at all other points as well and that they are equal to zero there.
The Wilson--Dirac operator may be considered to be a linear operator
in this space of fields, whose
action at $0<x_0<T$ is given by eq.~(5.1)
(elsewhere the target field is set to zero).
This is the lattice Dirac operator that was introduced by Sint
[\ref{SchroedingerFermion}].

The structure of eq.~(5.2) is such that $D$ 
satisfies the Ginsparg--Wilson relation (4.2) automatically 
(with $a$ replaced by $\abar$) if $\dirac{5}A$ is hermitian.
In the presence of the boundaries,
the Dirac operator must therefore be given by a different expression.
A formula that works out is
\equation{
  D={1\over\abar}\left\{
  1-\frac{1}{2}(\U+\Utilde)
  \right\},
  \enum
  \next{2ex}
  \U=A\left(A^{\dagger}A+c aP\right)^{-1/2}, 
  \qquad
  \Utilde=\dirac{5}U^{\dagger}\dirac{5},
  \enum
}
where $c\geq1$ is another tuneable parameter, whose optimal value will 
turn out to be close to $1+s$.
In eq.~(5.5), the boundary operator
\equation{
  P\psi(x)={1\over a}\left\{
  \delta_{x_0a}P_{-}\psi(x)|_{x_0=a}
  +\delta_{x_0\,T-a}P_{+}\psi(x)|_{x_0=T-a}
  \right\}
  \enum
}
is the four-dimensional version 
of the operator $P$ previously encountered,
while $A$ is again given by eq.~(5.3), where $\Dw$ is now the
Wilson--Dirac operator in the presence of the boundaries.

The merits of the definition (5.4),(5.5) will be discussed
in detail, but before this  
it may be helpful to note that the operator $D$ 
reduces to the Wilson--Dirac operator
in the one-dimensional theory if $s=0$ and $c=1$.
The results reported in sect.~4 actually imply that 
$A^{\dagger}A+caP=1$ in this case.

\subsection 5.2 Lattice symmetries, hermiticity and spectral bounds

It is not difficult to check that the Dirac operator $D$ 
transforms like the Wilson--Dirac operator
under cubic rotations, parity, 
time-reflections and charge conjugation. The latter interchanges
$\U$ with $\Utilde$, and having the sum of these two operators
in eq.~(5.4) also ensures that $\dirac{5}D$ is hermitian.

Another implication of the form (5.5) is the bound
\equation{
   \|\U\|=\|\Utilde\|\leq1.
   \enum
}
To show this, it suffices to note that
\equation{
  \|\U\psi\|^2=\left(\chi,A^{\dagger}A\chi\right)\leq
  \left(\chi,(A^{\dagger}A+caP)\chi\right)=\|\psi\|^2,
  \enum
}
for any quark field $\psi$, where 
$\chi=(A^{\dagger}A+caP)^{-1/2}\psi$.
The spectrum of $\abar D$ is thus
contained in the unit disk in the complex plane centred at $1$.
However, one should not expect the spectrum to be 
on the unit circle, as is the case for Dirac operators satisfying
the Ginsparg--Wilson relation.

In the present framework, the natural choice of the 
massive Dirac operator is [\ref{FerencReview}]
\equation{
  \Dm=(1-\frac{1}{2}\abar m)D+m.
  \enum
}
When the bare quark mass $m$ 
is in the range $0\leq m\leq 2/\abar$, as will be assumed in the following,
the spectrum of this operator is separated from the origin by a distance 
of at least $m$.
The middle term in the expansion
\equation{
  \left(\dirac{5}\Dm\right)^2=m^2+m(1-\frac{1}{2}\abar m)(D^{\dagger}+D)
  +(1-\frac{1}{2}\abar m)^2 \left(\dirac{5}D\right)^2
  \enum
}
is in fact non-negative
and the eigenvalues $\lambda_n$ of $\dirac{5}D_m$
are therefore bounded by
\equation{
  \lambda_n^2\geq m^2+(1-\frac{1}{2}\abar m)^2\mu^2,
  \enum
}
where $\mu$ denotes the spectral gap at $m=0$. 
This bound coincides with the spectral bound (2.8) 
in the continuum theory, up to corrections of order $am$.

An important consequence of these results is that
the determinant $\det\Dm$ is real and
non-zero at all quark masses $m>0$. 
Actually, $\det\Dm$ must be positive at these masses,
because this is trivially the case at $m=2/\abar$ and because
the determinant is a continuous function of $m$, on any finite lattice.

\subsection 5.3 Locality

In position space, the Dirac operator 
is represented by a kernel $D(x,y)$ through
\equation{
  D\psi(x)=a^4\sum_{y_0=a}^{T-a}\sum_{\mib y}D(x,y)\psi(y),
  \qquad 0<x_0<T.
  \enum
}
Locality requires that the bound
\equation{
  a^5\|D(x,y)\|\leq C\rme^{-\kappa\|x-y\|/a}
  \enum
}
holds for some constants $C$ and $\kappa>0$ that do not depend
on $a$.
Moreover, up to such exponentially small tails,
$D(x,y)$ should be locally constructed 
and only depend on the gauge field variables in the 
vicinity of $x$ and $y$. 

In infinite volume, a rigorous proof of the locality 
of the Neuberger--Dirac operator can be given
if the gauge field is not too rough on the scale of the lattice
spacing [\ref{Locality}]. Further studies then
suggest that locality holds under far more 
general conditions, including those typically 
encountered in numerical lattice QCD at lattice spacings
$a\leq0.1$ fm
[\ref{Locality},\ref{MobilityGap}].

The proof presented in ref.~[\ref{Locality}] is based
on an expansion of the inverse square root in eq.~(5.2)
in Legendre polynomials.
The expansion converges rapidly if 
$A^{\dagger}A\geq\alpha$ for some $\alpha>0$,
and the locality of the Dirac operator then follows immediately.
In the case of the operator (5.5), 
the Legendre expansion similarly 
links its locality properties to the existence of a
non-zero lower bound on $A^{\dagger}A+caP$.

As shown in appendix A, the spectral gap of 
$A^{\dagger}A+caP$ is not smaller than that of 
$A^{\dagger}A$ on the infinite lattice, independently of how precisely
the gauge field is extended from the time slice $0\leq x_0\leq T$ to 
all times. 
The field may be extended through time reflections
at the planes $x_0=0\mod T$, for example, which is a
good choice in the present context, since the smoothness
properties of the field (if any) are preserved.
In particular, the estimates of ref.~[\ref{Locality}]
then immediately imply the existence of a spectral gap 
if the gauge field at $0\leq x_0\leq T$ is sufficiently smooth
on the scale of the lattice spacing.

Presumably the locality properties of the Dirac operator (5.4)
are thus as good as those of the Neuberger--Dirac operator 
on lattices with periodic boundary conditions,
also when the rigorous
arguments of ref.~[\ref{Locality}] do not apply. However, some numerical
studies may still be required to confirm this in the cases of
interest.

\subsection 5.4 Chiral symmetry

At a distance $d$ from the boundaries, the kernel $D(x,y)$ 
of the Dirac operator (5.4)
coincides with the kernel of the Neuberger--Dirac operator
on the infinite lattice, up to terms that decrease exponentially
like $\rme^{-\kappa d/a}$.
From the expansion in Legendre polynomials 
mentioned in the previous subsection, for example,
this property is evident,
taking into account the fact that the operator $P$ 
is supported at the boundaries of the lattice.
The separation of bulk and boundary
terms is actually a direct
consequence of the locality of the operators involved
(see appendix B).

The remark has two important implications. First of all, it shows
that the Schr\"o\-dinger functional constructed here
probes the right theory, i.e.~the one where the lattice
Dirac operator in the absence of the boundaries 
is equal to the Neuberger--Dirac operator.
Secondly, it follows that 
\equation{
  \dirac{5}D+D\dirac{5}=\abar D\dirac{5}D+\Delta_B,
  \enum
}
where $\Delta_B$ is a local operator with kernel $\Delta_B(x,y)$ 
supported
in the vicinity of the boundaries 
(up to the usual exponentially small tails). 
Correlation functions of local fields at 
physical distances from the boundaries
thus satisfy the same chiral Ward identities 
as they do on lattices with 
periodic boundary conditions, for example.

Starting from the definition (5.4),(5.5) of the Dirac operator,
or from the formulae in appendix B, the operator $\Delta_B$
can be worked out explicitly. 
One may hope to find $\Delta_B=\dirac{5}P$, as is the 
case in one dimension, 
but the expressions that are obtained
are complicated and not very illuminating.

\subsection 5.5 Boundary fields and\/ {\rm O($a$)} improvement

A possible lattice representation of 
the boundary fields (2.5) at time $x_0=0$ is
\equation{
   \zeta({\mib x})=U(x,0)|_{x_0=0}P_{-}\psi(x)|_{x_0=a},
   \enum
   \next{2.5ex}
   \zetabar({\mib x})=\psibar(x)|_{x_0=a}P_{+}U(x,0)^{-1}|_{x_0=0},
   \enum
}
where $U(x,\mu)$ denotes the link variable at the point $x$ in direction
$\mu$. This definition coincides with the one commonly adopted in 
lattice QCD with Wilson quarks [\ref{SchroedingerFermion}].
It should be noted, however, that the normalization of these
fields depends on the details of the lattice regularization and 
may not be the canonical one (cf.~subsect.~5.6).

Formulations of lattice QCD with Ginsparg--Wilson quarks are
automatically on-shell O($a$)-improved [\ref{SW},\ref{OaImp}]. 
In the presence of the boundaries, this is no longer the case,
but the theory can be improved by including a few 
O($a$) boundary counterterms in the lattice action. 
The list of terms that must be added 
was determined in ref.~[\ref{OaImp}]. 
One of the counterterms amounts to a modification
of the lattice Dirac operator, while all others are either pure
gauge terms or reduce to contact terms and $1+\rmO(am)$ 
renormalization factors.

These other counterterms
can be implemented as in the standard lattice theory
[\ref{SchroedingerFermion},\ref{OaImp}].
The situation is a bit more tricky in the case of 
the counterterm that modifies the lattice Dirac operator,
because some of the desirable properties of the latter
may be compromised [the spectral bound (5.11), for example].

The counterterm at the boundary $x_0=0$ is usually taken to be 
a straightforward lattice implementation of the boundary action
\equation{
  a\int_{x_0=0}\rmd^3{\mib x}
  \left\{
  \psibar(x)P_{+}D_0\psi(x)+\psibar(x)\lvec{D}_0P_{-}\psi(x)
  \right\}.
  \enum
}
The precise choices that one makes do not matter, since
the counterterm is uniquely determined by its symmetries and dimension,
up to redundant terms and corrections of higher order in $a$.
In the present context, the O($a$) improvement 
can therefore also be achieved
by tuning the coefficient $c$ on which
the Dirac operator (5.4),(5.5) depends.
Changes of this coefficient actually amount to a modification
of the operator in the vicinity of the boundaries by a local term
with the correct symmetries. The properties
of the Dirac operator discussed in the previous subsections 
are then preserved.

\vskip0.2cm minus 0.2cm

\subsection 5.6 Free-quark theory

In the absence of the gauge field, it is possible to check
explicitly that the lattice theory has the 
correct continuum limit and that the O($a$) improvement
works out in the way described in the previous subsection.

The operator under the square root in eq.~(5.5) 
assumes the form
\equation{
  A^{\dagger}A+caP=
  (1+s)^2+sa^2\sum_{\mu}\drvstar{\mu}\drv{\mu}+
  \frac{1}{2}a^4\sum_{\mu<\nu}\drvstar{\mu}\drv{\mu}\drvstar{\nu}\drv{\nu}
  +(c-1)aP
  \enum
}
in the free theory. As explained in subsect.~5.1,
the operator acts on quark fields in the 
presence of the boundaries.
Its eigenfunctions at $c=1$, for example,
are given by
\equation{
  \sin(p_0x_0)\,\rme^{i{\mib px}}, 
  \qquad p_0={n\pi\over T}, \qquad n=1,2,\ldots,T/a-1.
  \enum
}
Since the associated eigenvalues 
are bounded from below by $(1-|s|)^2$ if $|s|\leq1$,
the locality of the 
free Dirac operator is guaranteed at all $|s|<1$ and
$c\geq1$.

In the time-momentum representation,
the kernel $D(x,y)$ of the Dirac operator
can be worked out analytically to some extent.
The quark propagator, on the other hand, 
may be difficult to obtain in closed form, 
but it can be computed numerically
on lattices with hundreds of points in each direction,
using established techniques
(see ref.~[\ref{NumTech}], for example).
This allows the lattice propagator to be compared
with the continuum propagator in a large range of lattice spacings,
spatial momenta and quark masses.

Many checks were then performed, including the following three:

\vskip1ex\noindent
(1) It was verified that the quark propagator 
$\left\langle\psi(x)\psibar(y)\right\rangle$ at non-zero
distances from the boundary as well as the 
boundary-to-bulk propagator 
$\left\langle\zeta({\mib x})\psibar(y)\right\rangle$
have the correct continuum limit 
(explicit expressions for the continuum propagator can be found in
ref.~[\ref{FreeQuark}]).

\vskip1ex\noindent
(2) The full spectrum of $(\dirac{5}D)^2$ was computed and compared
with the spectrum in the continuum theory. In particular, 
the presence of any additional modes 
that would survive in the continuum limit
could be excluded in this way.

\vskip1ex\noindent
(3) The approach of the propagator to the continuum limit was studied and it
was shown that the lattice effects of order $a$ can be cancelled by tuning 
the parameter $c$ of the lattice Dirac operator and
the field normalization factors.

\vskip1ex\noindent
The O($a$) improvement is achieved for values of $c$ close to $1+s$,
and the improved, canonically normalized boundary field 
is given by $Z(1+bam)\zeta$ where $Z\simeq 1-s/4$ and
$b\simeq-3/4$. These normalization factors will be rarely needed,
but it is reassuring to note that they are in a range that is 
not uncommon for such factors. Moreover,
the observed convergence of the propagators to the 
continuum limit is quite similar to the one 
seen in the standard Wilson 
theory [\ref{FreeQuark}].

\section 6. Concluding remarks

Lattice Dirac operators that satisfy the Ginsparg--Wilson 
relation have little in common with 
the difference operators that are obtained
by ``discretizing'' the classical Dirac equation.
Classical concepts can in fact be
rather misleading in lattice field theory.
In particular, boundary conditions 
cannot simply be imposed on the lattice fields
but arise dynamically in the continuum limit.

Universality considerations gain additional importance
in this context, since they apply to any local theory,
independently of how 
complicated it may be.
The definition of the Schr\"odinger functional 
proposed in this paper heavily builds on such arguments.
It is clearly 
not the only possible construction, but the proposed 
formulation works out and has many attractive features.

In lattice QCD with Ginsparg--Wilson quarks,
the application of domain decomposition methods [\ref{SchwarzHMC}]
appeared to be excluded so far,
because it was not clear how to restrict the lattice
Dirac operator to blocks of lattice points.
Following the lines of sect.~5, it is now 
straightforward to come up with viable expressions
for the block Dirac operators. 
The boundary operator $P$, for example, may be fixed by
requiring eq.~(A.1) to remain valid when $R$ is 
set to the block restriction operator. 
The locality 
of the block Dirac operators 
as well as a number of other desirable properties
are then guaranteed.

\vskip1ex
I am indebted to Stefan Sint for helpful discussions on
the Schr\"odinger functional
with chirally rotated boundary conditions
and to Peter Weisz for a critical reading of the paper.

\appendix A. Spectral bound on $\mib{A^{\dagger}A+caP}$

As explained in subsect.~5.3, the locality 
of the Dirac operator (5.4) can be proved rigorously if
$A^{\dagger}A+caP\geq\alpha$ for some $\alpha>0$.
In this appendix, it is shown that 
such a bound can be obtained by relating $A^{\dagger}A+caP$ to
the operator $A^{\dagger}A$ on the infinite lattice.

The standard setup of the lattice Schr\"odinger functional assumes
the gauge field variables $U(x,\mu)$ to be defined 
at all times $0\leq x_0<T$ if $\mu=0$ and additionally at $x_0=T$
if $\mu=1,2,3$. Evidently, a given field can always be extended to 
all times by setting the so far undefined link variables
to unity, for example. 

Exactly which extension is chosen will
not matter in the following. Once a definite prescription is adopted,
the Wilson--Dirac operator becomes a well-defined, bounded linear
operator in the space of all square-summable quark fields on 
the infinite lattice. The associated operator $A$ [eq.~(5.3)]
is denoted by $\Ahat$ in order to avoid any confusion with the operator
$A$ that appears in eq.~(5.5).

A straightforward calculation, similar to the one that leads to eq.~(4.5), 
now shows that
\equation{
  R\left(A^{\dagger}A+aP\right)R=R\hat{A}^{\dagger}\hat{A}R,
  \enum
}
where $R$ projects the quark fields to the physical subspace
in the presence of the boundaries. Explicitly,
\equation{
  R\psi(x)=\cases{\psi(x) & if $0<x_0<T$, \cr
                  \noalign{\vskip 1ex}
                  0       & otherwise,\cr}
  \enum
}
for any quark field $\psi$ on the infinite lattice. Note that
the product on the left-hand side of eq.~(A.1) is perfectly
well-defined, since $R$ projects the fields to the physical subspace 
before the operator in brackets (which can only act on quark 
fields in this space)
is applied.

Quark fields
in the physical subspace satisfy $R\psi=\psi$, and 
from eq.~(A.1) it then follows that
\equation{
   (\psi,(A^{\dagger}A+caP)\psi)=
   (\psi,\hat{A}^{\dagger}\hat{A}\psi)+
   (\psi,(c-1)aP\psi)\geq(\psi,\hat{A}^{\dagger}\hat{A}\psi),
   \enum
}
where $c\geq1$ was used and also the fact that $aP$ is a projector.
Since (A.3) holds for all quark fields in the physical subspace,
this shows that 
the operator $A^{\dagger}A+caP$ is bounded from 
below by the spectral gap of $\hat{A}^{\dagger}\hat{A}$,
as asserted in subsect.~5.3.

\appendix B. Separation of bulk and boundary terms

Equation (A.1) relates the theory in the presence of the boundaries
to the one on the infinite lattice. The goal in the following lines
is to work out the relation between the corresponding
Dirac operators. Along the way, a separation of bulk and boundary
terms is achieved, which allows the 
position-space kernels of the operators to be compared with each other.

The starting point is the identity
\equation{
   R(A^{\dagger}A+caP)R+(1-R)\Ahat^{\dagger}\Ahat(1-R)=
   \Ahat^{\dagger}\Ahat+\Bhat,
   \enum
}
in which 
\equation{
   \Bhat=(c-1)RaPR-
         (1-R)\Ahat^{\dagger}\Ahat R-
         R\Ahat^{\dagger}\Ahat(1-R)
   \enum
}
denotes an operator supported at the boundaries. 
Since $A^{\dagger}A+caP$ operates in the physical subspace,
and since $R$ is a projector, eq.~(B.1) implies
\equation{
    R(A^{\dagger}A+caP)^{-1/2}R=
    R(\Ahat^{\dagger}\Ahat+\Bhat)^{-1/2}R.
    \enum
}
Using a well-known integral representation,
the operator on the right-hand side of this equation can be
written in the form
\equation{
   (\Ahat^{\dagger}\Ahat+\Bhat)^{-1/2}=
   \noenum
   \next{2ex}
   \qquad
   (\Ahat^{\dagger}\Ahat)^{-1/2}-
   \int_{-\infty}^{\infty}{\rmd t\over\pi}\,
   (\Ahat^{\dagger}\Ahat+t^2)^{-1}\Bhat
   (\Ahat^{\dagger}\Ahat+\Bhat+t^2)^{-1}.
   \enum
}
Taken together, these equations provide a representation of 
$(A^{\dagger}A+caP)^{-1/2}$ (and thus of the Dirac operator in 
the presence of the boundaries) in terms of the operator
$(\Ahat^{\dagger}\Ahat)^{-1/2}$ on the infinite lattice plus
another operator localized at the boundaries. Both 
$(\Ahat^{\dagger}\Ahat+t^2)^{-1}$ and
$(\Ahat^{\dagger}\Ahat+\Bhat+t^2)^{-1}$
are in fact expected to 
be local operators, and 
the last term in eq.~(B.4) is thus
supported in the vicinity of the boundaries
(up to exponentially small tails).

\beginbibliography

% Schroedinger functional in QCD

\bibitem{SchroedingerGauge}
M. L\"uscher, R. Narayanan, P. Weisz, U. Wolff,
Nucl. Phys. B384 (1992) 168

\bibitem{SchroedingerFermion}
S. Sint, 
Nucl. Phys. B421 (1994) 135;
{\it ibid.} B451 (1995) 416

\bibitem{LesHouches}
M. L\"uscher,
{\it Advanced lattice QCD},
in: Probing the
Standard Model of Particle Interactions (Les Houches 1997),
eds. R. Gupta et al.
(Elsevier, Amsterdam, 1999)

% Exact chiral symmetry on the lattice

% Ginsparg-Wilson relation

\bibitem{GinspargWilson}
P. H. Ginsparg, K. G. Wilson,
Phys. Rev. D25 (1982) 2649

% Classically perfect Dirac operator

\bibitem{Hasenfratz}
P. Hasenfratz,
Nucl. Phys. B (Proc. Suppl.) 63 (1998) 53;
Nucl. Phys. B525 (1998) 401

\bibitem{HLN}
P. Hasenfratz, V. Laliena, F. Niedermayer,
Phys. Lett. B427 (1998) 125

% Neuberger-Dirac operator

\bibitem{NeubergerDirac}
H. Neuberger,
Phys. Lett. B417 (1998) 141;
{\it ibid.}\/ B427 (1998) 353

% Exact chiral symmetry

\bibitem{ExactChSy}
M. L\"uscher,
Phys. Lett. B428 (1998) 342

% Orbifold construction of the Schroedinger functional 

\bibitem{OrbiTaniguchi}
Y. Taniguchi,
JHEP 0512 (2005) 037

\bibitem{OrbiSint}
S. Sint,
PoS LAT2005 (2005) 235

% Renormalization of quantum field theories in presence of boundaries

\bibitem{SchrodingerSymanzik}
K. Symanzik,
Nucl. Phys. B190 [FS3] (1981) 1

% Boundary critical phenomena

\bibitem{Diehl}
H. W. Diehl,
Int. J. Mod. Phys. B11 (1997) 3503

% Wilson's formulation of lattice QCD

\bibitem{Wilson}
K. G. Wilson, 
Phys. Rev. D10 (1974) 2445

% Massive Dirac operator

\bibitem{FerencReview}
F. Niedermayer,
Nucl. Phys. (Proc. Suppl.) 73 (1999) 105

% Locality of the Neuberger-Dirac operator

\bibitem{Locality}
P. Hern\'andez, K. Jansen, M. L\"uscher,
Nucl. Phys. B552 (1999) 363

\bibitem{MobilityGap}
M. Golterman, Y. Shamir, B. Svetitsky,
Phys. Rev. D72 (2005) 034501;
PoS LAT2005 (2005) 129

% O(a) improved lattice QCD

\bibitem{SW}
B. Sheikholeslami, R. Wohlert,
Nucl. Phys. B259 (1985) 572

\bibitem{OaImp}
M. L\"uscher, S. Sint, R. Sommer, P. Weisz,
Nucl. Phys. B478 (1996) 365

% Minmax polynomial

\bibitem{NumTech}
L. Giusti, C. Hoelbling, M. L\"uscher, H. Wittig,
Comput. Phys. Commun. 153 (2003) 31

% Free quark propagator

\bibitem{FreeQuark}
M. L\"uscher, P. Weisz,
Nucl. Phys. B479 (1996) 429

% Domain decomposition in LQCD

\bibitem{SchwarzHMC}
M. L\"uscher,
Comput. Phys. Commun. 156 (2004) 209;
{\it ibid.} 165 (2005) 199

\endbibliography

\bye